\documentstyle[epsfig,12pt,a4]{article}

\date{}                             

\begin{document}
        
\title{ON PAIR PRODUCTION OF SCALAR TOP QUARKS IN 
            $e^{+}e^{-}$ COLLISIONS AT ILC AND A POSSIBILITY 
	    OF THEIR MASS  RECONSTRUCTION}

\author{A.~Bartl$^{a,b}$, ~~ W.~Majerotto$^{c}$, \\
K.~M$\ddot o$nig$^{d}$, ~~ A.N.~Skachkova$^{e}$, 
~~  N.B.~Skachkov$^{e}$ }

\maketitle
\begin{center}
{\normalsize \it $^a$ University of Vienna, Faculty of Physics, 
1090 Vienna, Boltzmanngasse 5, Austria. \\
 $^b$ AHEP Group, Instituto de Fisica Corpuscular - C.S.I.C.,
  Universidad de Valencia, Edificio Institutos de Investigacion,
   Apt. 22085, E-46071 Valencia, Spain. \\
$^c$ Institute for High Energy Physics (HEPHY Vienna),
 Nikolsdorfergasse 18, A-1050 Vienna, Austria.\\
$^d$ DESY, Platanenallee 6,   D-15738 Zeuthen, Germany. \\
$^e$ JINR, Joliot-Curie 6, 141980 Dubna,  Moscow region, Russia. \\}
\end{center}

\bigskip
\begin{abstract}
\noindent 

    We study  pair production of scalar top quarks
    (stop, $\tilde t_{1}$) in  $e^{+}e^{-}$ collisions 
    with the subsequent decay of the top squarks
    into  $b$-quarks and 
    charginos  $\tilde t_{1} \to b \tilde \chi_{1}^{\pm}$. 
    We simulate this process  by using PYTHIA6.4
    for   the beam energy $2E_{b}=\sqrt {s} = 
    350, 400, 500, 800, 1000$ GeV.
    A set of criteria for physical variables 
    is proposed,  which provides good
    separation  of stop signal events from top 
    quark pair production  being  the main
    background.  These  criteria allow us  to
    reconstruct the mass of the  top squark 
    with an integrared luminosity of 1000 $fb^{-1}$
    provided  that the neutralino mass is known.
\end{abstract}


\section{Introduction.}

 ~~~~The scalar top quark, the bosonic
 partner of the top quark, has attracted much attention 
 as it is expected to be the lightest colored 
 supersymmetric (SUSY) 
 particle. $\tilde t_{L}$ and $\tilde t_{R}$, the
 supersymmetric partners of the
 left-handed and right-handed top quarks, 
 mix and the
 resulting two mass eigenstates $\tilde t_{1}$ and
 $\tilde t_{2}$, can	 
 have a large mass splitting. It is even possible
 that the lighter eigenstate $\tilde t_{1}$ could be 
 lighter than the top quark itself \cite{JEllis}.

  Searches for top squarks which 
  were performed at  LEP and Tevatron 
  \cite{D0}, have shown that the mass of the stop is 
  higher than  141 GeV for a mass difference between 
  the stop and the lightest neutralino of about 50-70 GeV.
  These searches   will continue 
  at  LHC and ILC 
 \cite{ILCRDR1}, \cite{ILCRDR2}. 
%
  
  In 
  the following we study the reaction 
  \footnote{
            More details about this study can be
           found in \cite{BMMSS-ee}.
	    The analogous analysis in the photon-photon channel 
	    was made in \cite{BMMSS-STOP}. }
\begin{equation}
      e^{+} + e^{-} \to \tilde t_{1} + \bar{\tilde t_{1}}~. 
\end{equation}
 
  Among the possible  $\tilde t_{1}$-decay channels 
 within the MSSM (see \cite{Bartl} for details),  
 we focus on the decay
 $\tilde t_{1} \to b \tilde \chi_{1}^{\pm}$
  followed by the two-body chargino decay 
 $\tilde \chi_{1}^{\pm} \to  \tilde \chi_{1}^{0} W^{\pm}$,
 where one of the W's decays hadronically, 
  $W \to q_{i}{\bar q_{j}} $, 
 and the other one leptonically, 
 $W \to \mu\nu_{\mu}$ \cite{Paris2004}
 \footnote{ The process
          $e^{+}e^{-} \to \tilde t_{1} + \bar{\tilde t_{1}}$
          with  the subsequent decay
          $\tilde t_{1} \to  c \tilde \chi_{1}^{0}$
	  was  considered
	 in \cite{H.Nowak2}.}.
  The final state of this signal process, shown in the left-hand
 plot of Fig.1, contains 
 two $b$-jets and two (or more) jets (originating  from the
 decay of  one W boson), a hard muon  plus a  neutrino 
 (from the decay of the other W) and two neutralinos:
 \begin{equation}
   e^{+}e^{-} \to \tilde t_{1} \bar{\tilde t_{1}} \to
  b\bar{b}\tilde\chi^{+}_{1}\tilde\chi^{-}_{1} \to
   b \bar{b}W^{+}W^{-}\tilde\chi^{0}_{1}\tilde\chi^{0}_{1} \to 
   b\bar{b}q_{i}\bar{q_{j}}\mu\nu_{\mu}\tilde
    \chi^{0}_{1}\tilde\chi^{0}_{1}~. 
 \end{equation}
 The main background process is top quark pair production 
 with the subsequent decay $t \to bW^{\pm}$ (for W's we 
 use the same decay channels as in the stop case): 
  \begin{equation}
    e^{+}e^{-} \to  t \bar{ t} \to  b \bar{b}W^{+}W^{-} \to 
   b\bar{b}q_{i}\bar{q_{j}}\mu\nu_{\mu}~.
 \end{equation}
 The only difference between the final states of  stop and top 
 production 
(shown in the right diagram
 of Fig.1) is that in
 stop pair production there are two neutralinos 
 which are undetectable.  Thus,  both processes 
  have
 the same signature: two $b$-jets, two jets from W decay and
 a  muon. 
 In the present paper we  consider only top pair 
 production as  background.
      \begin{figure}[!ht]
     \begin{center}
 \vskip -0.5 cm      
    \begin{tabular}{cc}
      \mbox{\epsfig{file=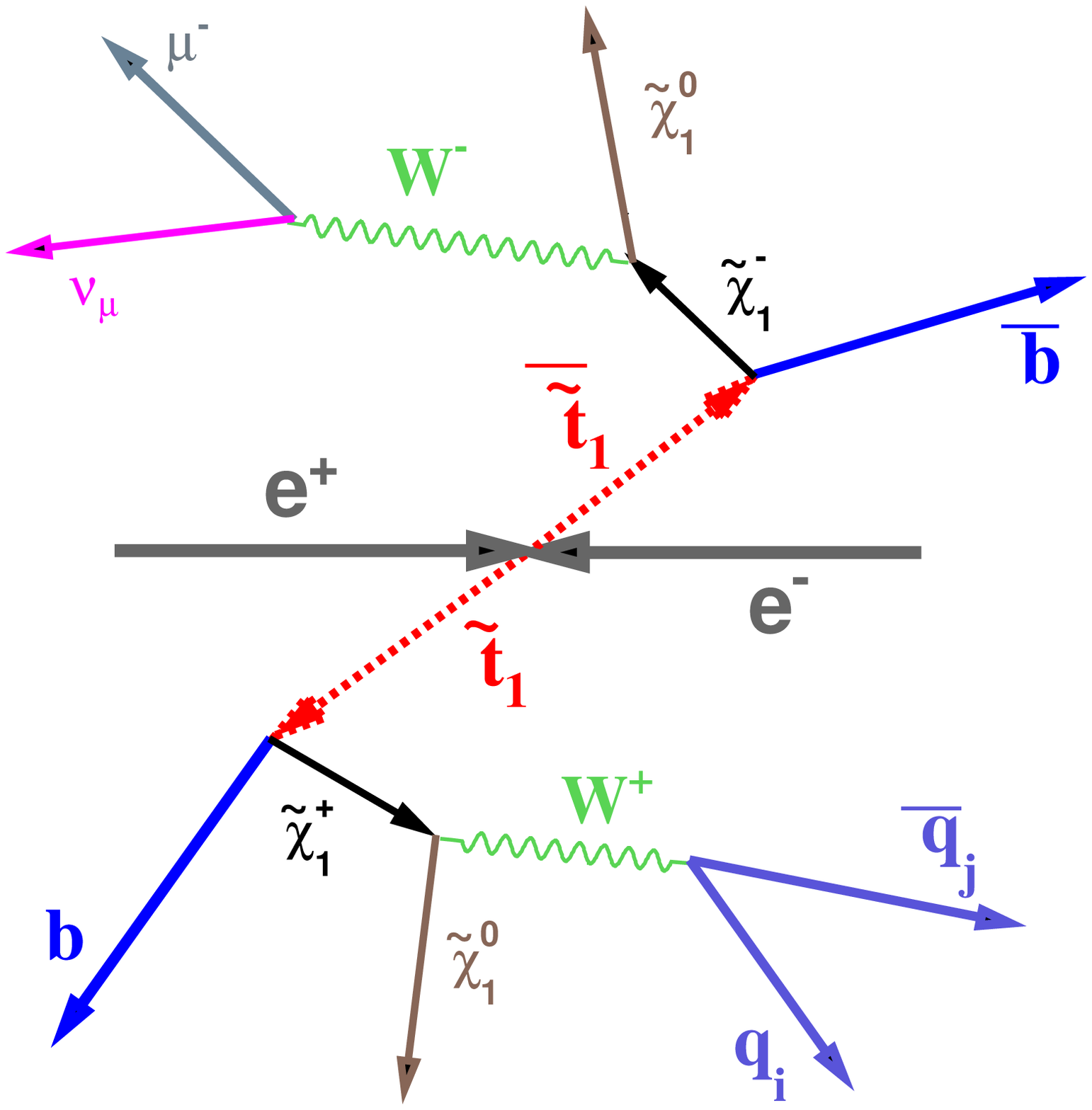,
             width=7.8cm, height=5.28cm}} 
     \mbox{\epsfig{file=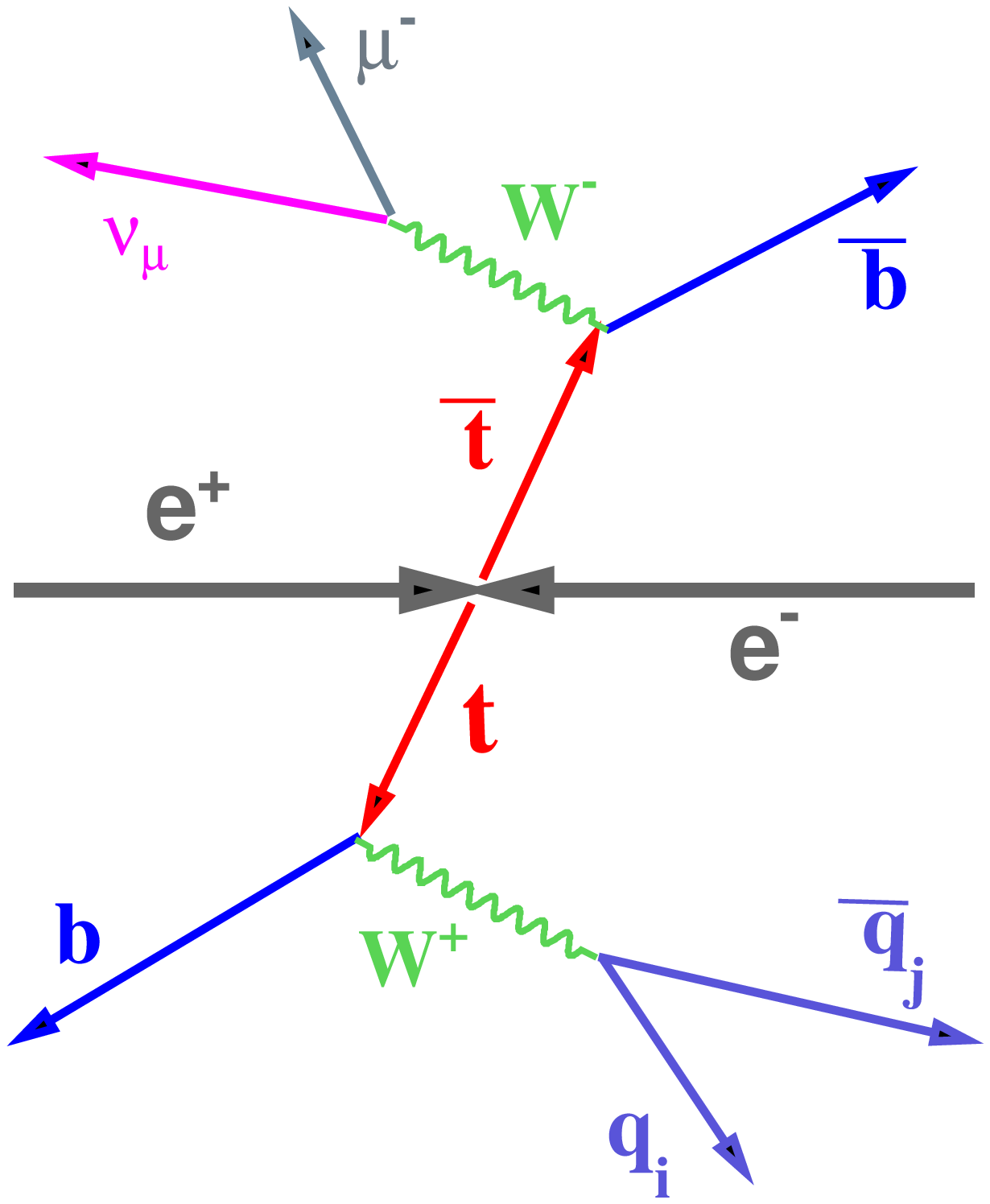,
                      width=7.8cm, height=5.28cm}} \\    
    \end{tabular}
     \caption{\it \small Left is the stop signal 
                         event diagram,  Right is 
			 the top background diagram. }    
     \end{center}  
  \vskip -0.5 cm           
     \end{figure}
    The analysis of the processes (2) and (3) is
 done on the basis of  Monte Carlo  samples of the
  corresponding events
 generated by the two programs PYTHIA6.4
  \cite{T. Sjostrand} and  CIRCE1
  \cite{T.Ohl}. 
   The program CIRCE1 
 is used for  the 
  parameterization of the beam spectra 
  involved in processes (1) and (3) to account
  for the effects of beamstrahlung.
    The energy of the  beams
  was chosen as 
   $2E_{b}=\sqrt {s}$ = 350, 400, 500, 800, 1000 GeV.
 
    In Section 2 we give the set of MSSM parameters used in 
  our study.

   In Section 3 we discuss  some  general characteristics 
  of the signal process
  $e^{+}e^{-} \to \tilde t_{1} \tilde {\bar{t}}_{1}$
  and the main background  $ e^{+}e^{-} \to t\bar{t}$.
  First we show the result of 
  taking into account 
  beam interaction effects  like  beamstrahlung and others
  which are simulated 
 with CIRCE1 \cite{T.Ohl}.
  The distributions for 
   a set of physical  variables, which
  are measurable in the  final state of the  process (2),
  are given for 
  the case of 
   a beam energy 
   $\sqrt {s}$ = 500 GeV  and 
  integrated luminosity of 1000 $fb^{-1}$.  These
   variables 
   include  hadronic jets  from W-decay
   as well as $b$-jets produced in stop decays.
  We always compare them in detail with 
  those of  top pair production. 
  
   In Section 4 we 
   propose  three cuts which allow
  a good separation of  signal stop events and 
  background  top events. The  number of stop 
  events that have passed  these  cuts and the
  values of the  corresponding  efficiencies are given.
  We show  that the invariant mass  
  of one $b$-jet and 
  the other two $non-b$-jets  (from W decay)  allows 
  one to reconstruct the mass of the scalar top quark
  provided that  the neutralino mass is known.
     The impact of the proposed  cuts on 
  the values of the cross sections of stop and
  top pair productions are shown together with
  the values of  signal-to-background ratios $(S/B)$.
  We have
  also considered 
  a cut which
  allows to separate a signal muon from W
  decay and  fake muons appearing 
  due to  decays of  other particles  in the
  same stop production event.    
   Section 5 contains some  conclusions.   
\section{ MSSM parameters and cross section.}
  ~~~~ The scalar top quark system is described 
  by the mass matrix 
   (in the $\tilde t_{L} - \tilde {t}_{R}$ basis)
  \cite{JEllis},   \cite{Gunion}
\begin{equation}    
   \left(\begin{array}{cc} M^2_{\tilde t_{LL}} & 
   M^2_{\tilde t_{LR}} \\
    
           M^2_{\tilde t_{RL}} & 
	   M^2_{\tilde t_{RR}} \end{array}\right) 
\end{equation}	   
with
\begin{equation}  
 M^2_{\tilde t_{LL}} = M^2_{\tilde Q} + 
 (\frac{1}{2} - \frac{2}{3} sin^2
 \Theta_W) cos2 \beta M^2_Z + M^2_t, 
\end{equation}
\begin{equation} 
M^2_{\tilde t_{RR}} = M^2_{\tilde U} +  \frac{2}{3} 
sin^2 \Theta_W cos2 \beta M^2_Z + M^2_t, 
\end{equation}
\begin{equation}  
M^2_{\tilde t_{RL}} = (M^2_{\tilde t_{LR}})^* =  M_t(A_t - \mu ^* cot \beta). 
\end{equation}
   The mass eigenvalues are given by
\begin{equation} 
 M^2_{\tilde t_{1,2}} = \frac{1}{2}\left[ (M^2_{\tilde t_{LL}} + 
  M^2_{\tilde t_{RR}}) \mp \sqrt{(M^2_{\tilde t_{LL}} + 
  M^2_{\tilde t_{RR}}) + 4 |M^2_{\tilde t_{LR}}|} \right]
\end{equation}
with the mixing angle
\begin{equation}
cos \theta_{\tilde t} = \frac {-M^2_{\tilde t_{LR}}} {\sqrt {
|M^2_{\tilde t_{LR}}|^2 + (M^2_{\tilde t_{1}} -  M^2_{\tilde t_{LL}} )^2}},
\end{equation}
\begin{equation}
sin \theta_{\tilde t} = \frac {M^2_{\tilde t_{LL}} - M^2_{\tilde t_{1}} } {\sqrt {
|M^2_{\tilde t_{LR}}|^2 + (M^2_{\tilde t_{1}} -  M^2_{\tilde t_{LL}} )^2}} .
\end{equation}
   In 
  the following we 
   will  consider 
  a particular 
  choice of the MSSM parameters that are defined, in the 
  notations of PYTHIA6.4,  in the following  way: \\
  
~~  $ M_{\widetilde{Q}} = 270$ GeV;
    ~~$ M_{\widetilde{U}} = 270$ GeV;  
   ~~$ A_t = -500$ GeV (top  trilinear coupling);  
    
~~~~~~~~~~~~ $ tan \beta = 5 $;~~ $\mu =  -370 $ GeV;  
  ~~ $ M_{1} = 80 $ GeV; 
  ~~ $ M_{2} = 160 $ GeV.\\    
%

  Note that in PYTHIA6.4 $M_{\widetilde{Q}} $  corresponds
  to $ M_{\widetilde{t}_L} $
  (left squark mass for the third generation) and 
  $M_{\widetilde{U}}$  corresponds to $M_{\widetilde{t}_R}$.
  These parameters give $M_{\widetilde{t}_1}=167.9$ GeV, 
  $M_{\chi^{+}_{1}}=159.2$ GeV 
  and  $M_{\chi^{0}_{1}}=80.9$ GeV. 
  This value of $ M_{\widetilde{t}_1}$ is rather close
  to the mass of the top quark
  $M_{top}=170.9 \pm 1.8$ GeV 
  \cite{Schieferdecker}.
  Therefore one expects a rather  large  contribution 
  from the top background, which means that the choice 
  of this value of the stop mass makes the analysis 
  most difficult. Finding a suitable set of 
  cuts separating stop and top events 
  is crucial.
 
   In general, the cross section for stop pair 
   production at a
   fixed energy depends on the mass of the 
   stop quark and the mixing angle
   $\theta_{\tilde t}$. Since the couplings
   of the $Z^0$ to the left and right
   components of the stop are different,
   the cross sections depend significantly
   on the beam polarizations 
   (see \cite{Bartl}, \cite{Paris2004},
   \cite{Gudi}). By choosing 
   appropriate
   longitudinal beam polarizations it is possible
   to enhance the cross sections. For
   example, for an electron beam with 90$\%$
   left polarization the cross section
    would be larger than the unpolarized
   cross section by approximately 40$\%$, for
    cos $\theta_{\tilde t}$ = - 0.81
   corresponding to the parameters given 
   above. If in addition the positron
   beam has 60$\%$ right polarization, then the
   cross section is enhanced by approximately
   a factor of 2 compared to the 
   unpolarized cross section. We note that
   a rather precise determination of the
   stop mixing angle $\theta_{\tilde t}$
   is possible by measuring the left-right 
   asymmetry. The cross section for
   top pair production has also a
   characteristic dependence on the beam
   polarizations \cite{Gudi}. For example, 
   the polarization of both beams
   leads to an increase of the cross 
   section by about a factor of 1.5. 
\section{Distributions of kinematic variables in 
         stop and top production.}
 ~~~~  In this Section we present some plots of 
  distributions for  different  physical variables
  based on $5 \cdot 10^{4}$ stop pair production 
  events  generated by PYTHIA6.4   and CIRCE1  
  weighted 
  with the electron-positron
  luminosity.  Analogous plots are also given for
  $ 10^{6}$  generated  background top events. 
  
  The ILC is a 200-500 GeV center-of-mass high luminosity
  (a peak luminosity of $\sim 2\cdot 10 ^{34} cm^{-2}s^{-1}$)
  linear electron-positron collider  with a possible upgrade 
  to 1~TeV 
  in the second phase.  
  According to \cite {ILCRDR1} 
 a total luminosity 
 of 500 fb$^{-1}$  
  is foreseen within  the first four years of 
  operation  and 1000 fb$^{-1}$ 
  during the first phase of operation
  at  500 GeV.  A first 
  run 
  at  $\sqrt {s}=500$ GeV 
  will get a first measurement of the particle masses to optimize
  the threshold scan \cite{ILCRDR2}.
  
   Fig.2 {\bf a)} demonstrates the total 
  energy spectrum of the electron and positron 
  beams, which is expected  at 
  $\sqrt {s}=500$ GeV  after taking into
  account beamstrahlung 
  and other beam interaction
  effects (see, for instance \cite{DShulte}).
  Fig.2 {\bf b)} shows the correlations of the
  beam fractions
  $y_i=E^{i}/E^{i}_{beam}$ (i=$e^+,e^-$)
  of the colliding  electron and positron beams.		    
 \begin{figure}[!ht]
     \begin{center}
 \vskip -0.5 cm     
    \begin{tabular}{cc}
     \mbox{a) \epsfig{file = 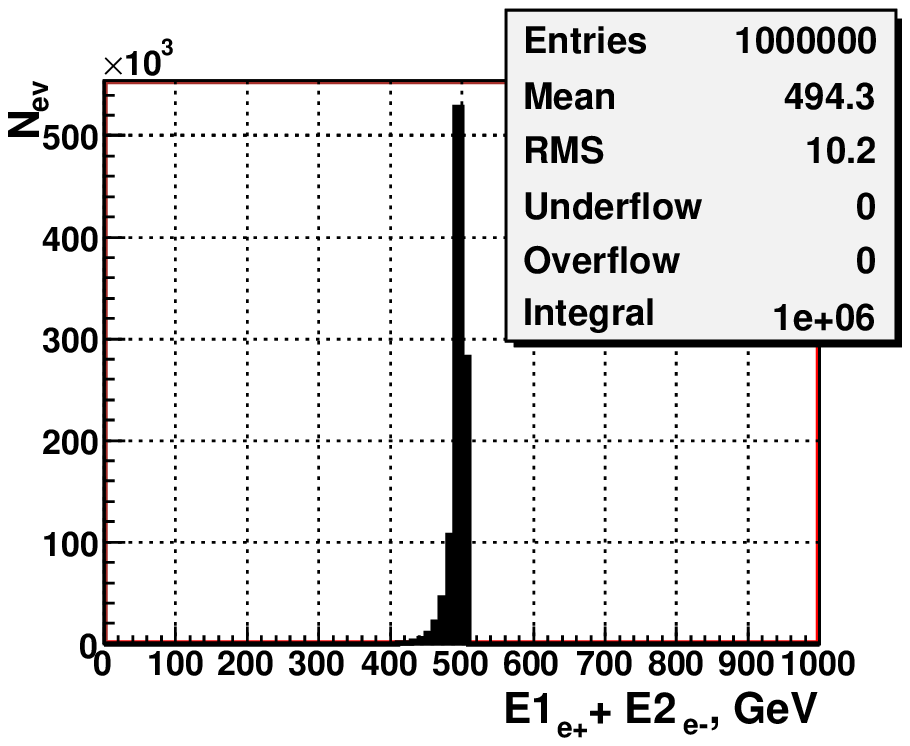,
                      width=7.2cm, height=5.2cm}}      
     \mbox{b) \epsfig{file = 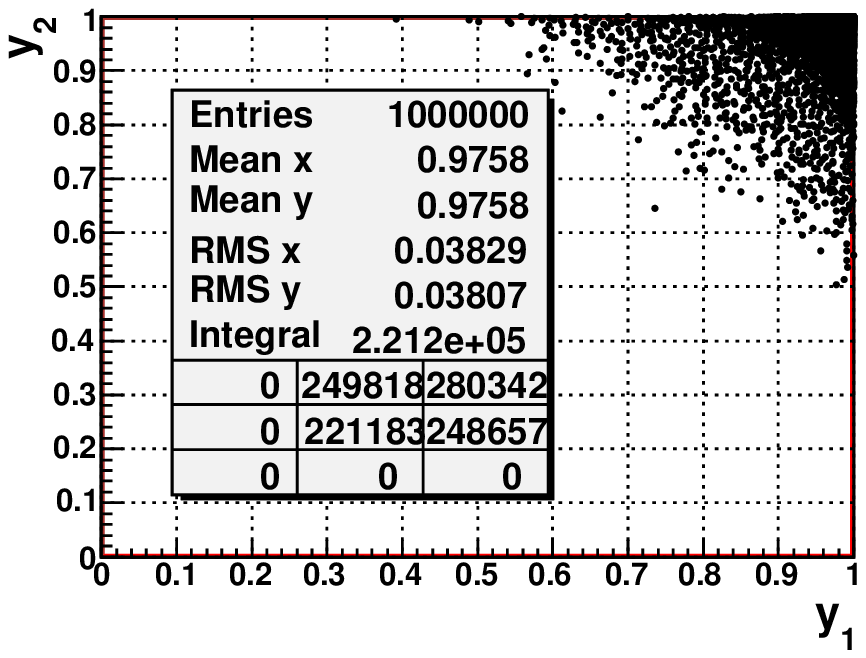,
                      width=7.2cm, height=5.2cm}} \\
    \end{tabular}
     \caption{\small \it {\bf a)} total beam energy spectrum,
     {\bf b)} beam fractions correlations }     
     \end{center}   
 \vskip -0.5 cm           
     \end{figure}
     
  To find the jets we used the subroutine PYCLUS of PYTHIA 
  with the distance measure used in the "Durham algorithm".
  The parameters of this jetfinder  are  chosen such 
  that the number of jets is exactly four. Technically, 
  $b$-jets are defined as jets that contain at 
  least one B-hadron. Their decay  may be identified 
  by the presence of a secondary vertex \cite {Haw}.
       
    Let us also mention that separate 
   stop and top event samples were generated without and 
   with cuts. This shows the effect of the cuts.    
%
\subsection{  Global  jet  variables.} 
%
 ~~~~    In this  subsection  we present some plots 
   that were obtained without cuts.  
   In addition to  the two $b$-jets there are also,  
   according to the decay chain (2), two jets due to the decay 
   of one W boson  into two quarks $W \to q_{i} + \bar q_{j}$
   (see Fig.1).  It is well known that the characteristics of
   individual jets depend on the choice of
   the jet finder and its parameters.
   This dependence may result in some  redistribution of
  energy between  jets.  To diminish
   these  effects
   we  use 
   three   physical variables which characterise the 
   physical system composed of  $b$-jets and the jets
   produced by quarks in
   the W decay. The simulation has
   shown that  
   the effect of energy redistribution between jets 
  is partially compensated in multijet
  systems.

   For this reason we consider first the invariant mass 
   of all four jets produced in the events
   which is the modulus of the vectorial sum  
   of the 4-momenta  $P^{j}_{jet}$
   \footnote{  $ P =(P_{0},  \bf P)$, $P_{0}=E_{p}$.
   We denote ${\bf{P}}^{j}_{jet}$
    as the 3-momentum   of the jet $j$: 
    $({\bf{P}}^{j})$$_{i}=(P$$^{j}$)$_{i}$,
   ($i= 1,2,3$).  
    The sum in (11)  is over all  four jets  ($j=1,2,3,4$). }
\begin{equation}    
   M_{inv}(All jets)=\sqrt{(\Sigma_{j=1,2,3,4}P^{j}_{jet})^{2}}.
\end{equation}     
    
%
%
%
   The distribution of this invariant mass is shown in Fig.3. 
  Plot {\bf a)} shows the results  for   stop pair 
  production 
    while the plot {\bf b)}  
  is for  top pair production. 
%
    In Fig.3 and the following figures the vertical axis
    shows the number 
   of stop and top events 
    that may be expected 
   for the integrated luminosity of 1000 $fb^{-1}$.
   Taking the integral of the distributions 
  one can get  the total  number  of events expected
   for the taken integral luminosity.
   These numbers   are shown as  "Integral" values 
   in the Figures.  One sees  that for the
   chosen luminosity 
   the number of the produced  background top
   events (35 930) is about 15 times higher than
   the number of stop events (2373). 
  It gives 
  a signal--to--background  ratio  $S/B=0.066$.

\begin{figure}[!ht]
     \begin{center}
\vskip -0.5cm       
   \mbox{a) \epsfig{file =   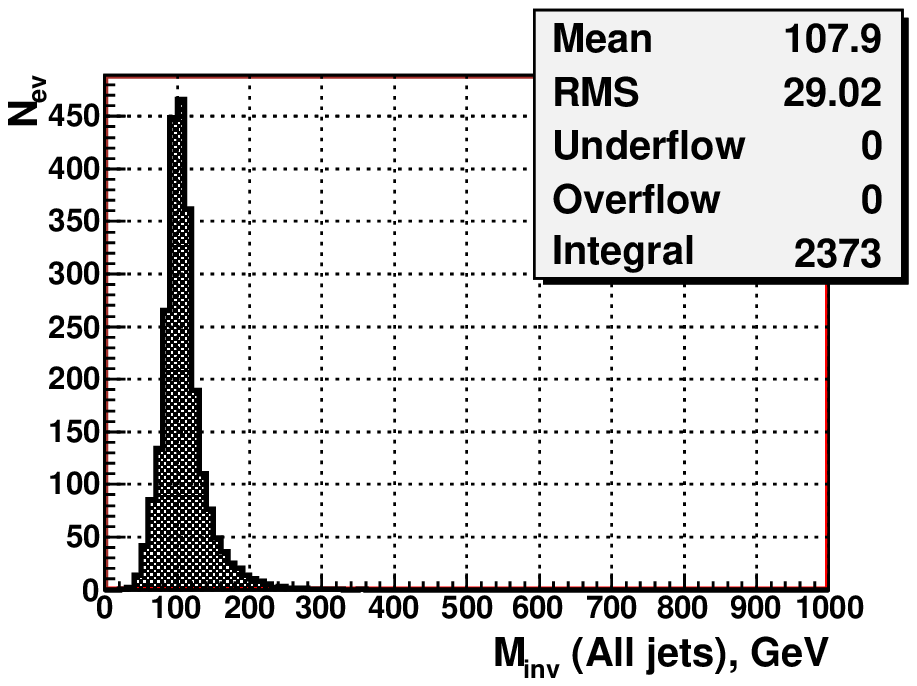,
                      width=7.2cm, height=5.2cm}}  
   \mbox{b) \epsfig{file = 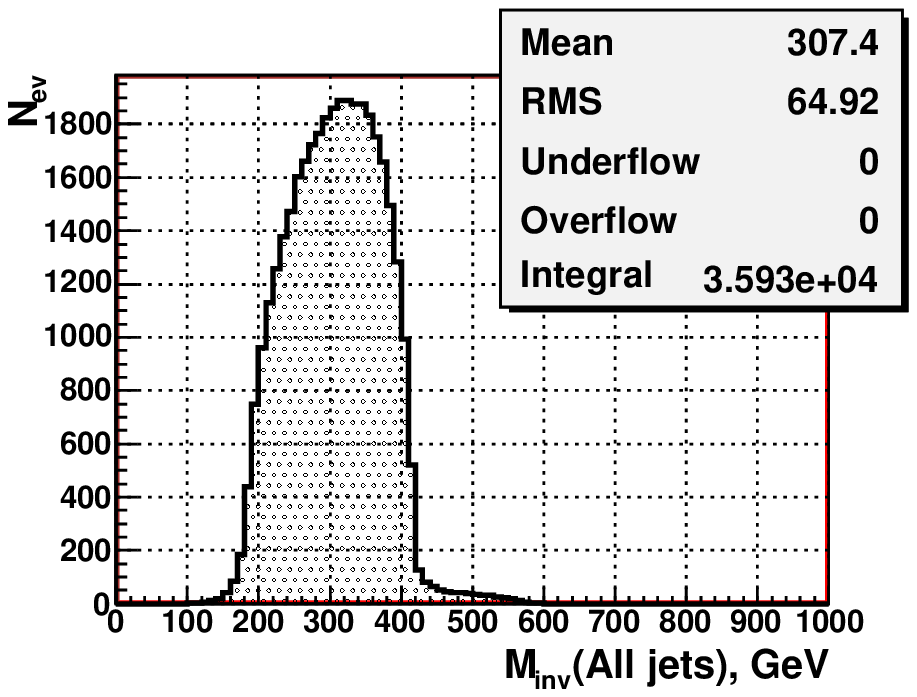,
                      width=7.2cm, height=5.2cm}}  \\
     \caption{\small \it Distribution 
                 of events versus the  reconstructed 
		 invariant mass of all jets $M_{inv}(All jets)$.
	       	 {\bf a)} stop pair production, 
	     {\bf b)} top pair production.}		 
     \end{center}    
\vskip -0.5cm           
     \end{figure}

  Plot  {\bf a)} of Fig.3  shows that  the  mean value 
  of $M_{inv}(All jets)$ is about three  times  lower 
  in stop production   than in top case, plot  {\bf b)}
  of Fig.3. Therefore, a rather soft cut like 
  $M_{inv}(All jets ) \leq 160$ Gev,  will
  strongly  suppress  the  top background and save the bulk
  of signal stop events.           
     Another useful variable that can also be used for 
  the separation of  the  signal  and the background is the "missing"
  mass  (we use  $\sqrt {s}=500$ GeV)
 \begin{equation}  
  M_{miss} =   \sqrt{ (\sqrt {s_{ee}} -
   (\Sigma_{j=1,2,3,4}E^{j}_{jet}+E_{\mu}))^{2}-
   (\Sigma_{j=1,2,3,4} {\bf P}_{jet}^{j}+\bf P_{\mu})^{2}}.
\end{equation} 
  This variable takes  into account  the contribution of 
  those particles that cannot be registered in the detector 
  (neutrinos and neutralinos). 
  The distributions of this  invariant "missing" mass 
  are given in Fig.4.   Plot {\bf a)} shows the results for  stop
  pair  production,  while the plot {\bf b)} is for 
  top pair production.  As seen from these plots, the cut 
  $M_{miss} \ge 250$~GeV will also allow us to get  rid of 
  most 
  of  the background events.

\begin{figure}[!ht]
     \begin{center}
\vskip -0.5cm             
\mbox{a) \epsfig{file =        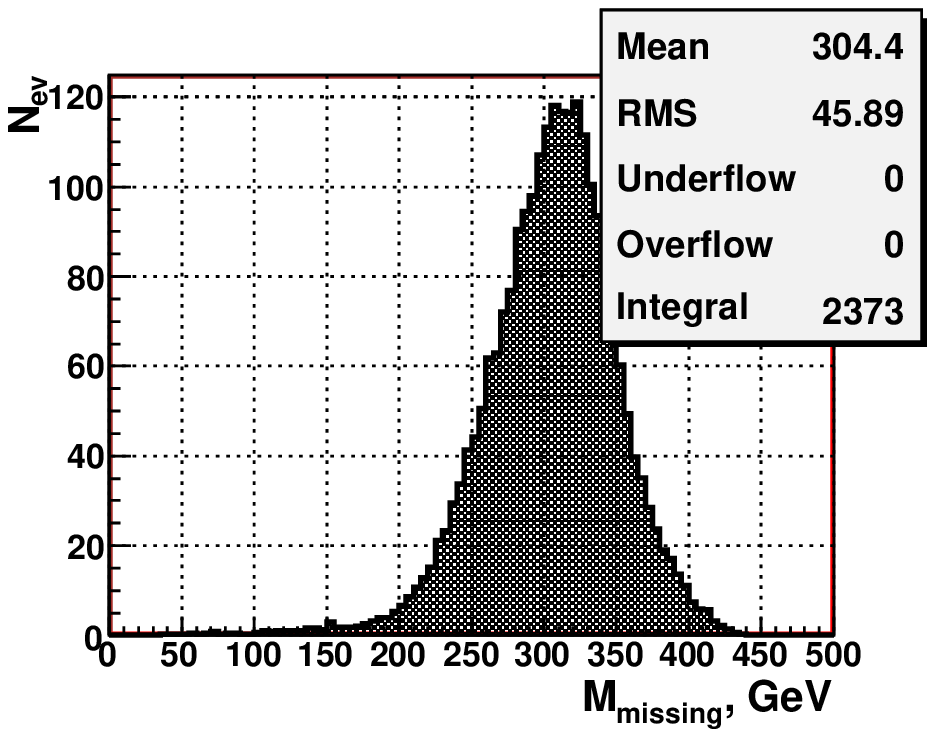,
                       width=7.2cm, height=5.2cm}}      
   \mbox{b) \epsfig{file = 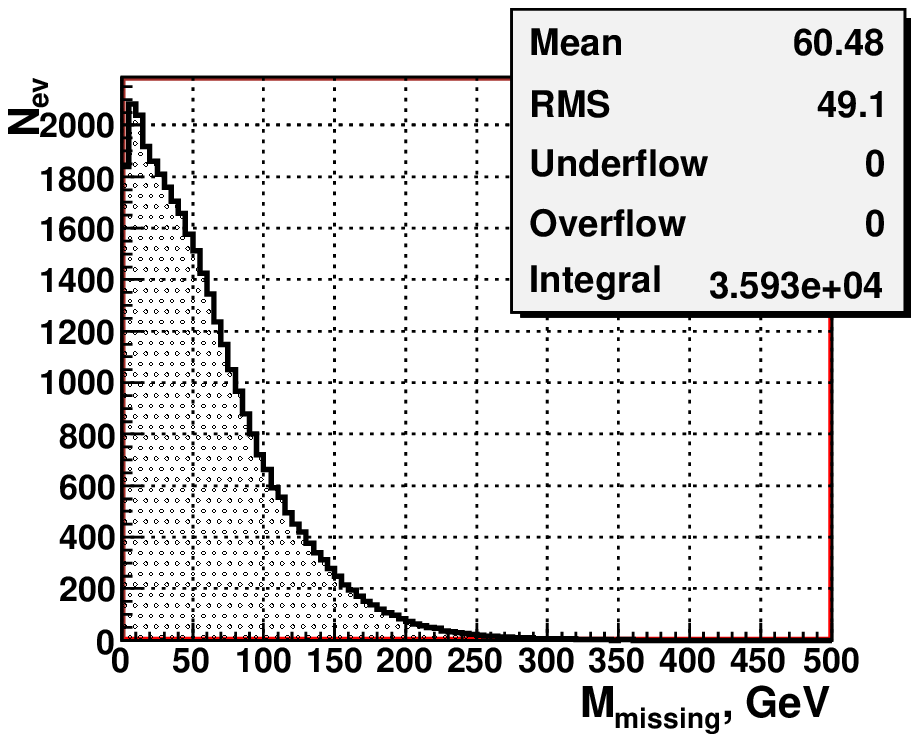,
                       width=7.2cm,  height=5.2cm}}      
     \caption{\small \it Distribution of number 
                 of events versus the missing mass variable.
		  {\bf a)} stop pair production, 
	      {\bf b)} top pair  production.}
     \end{center}   
\vskip -0.5cm                  
     \end{figure}

    ~~~~     The third variable of interest is the   
   invariant 
    mass $M_{inv}$($b$-jet,$JETS_{W}$)
\begin{eqnarray}
   M_{inv}(b-jet,JETS_{W}) \equiv
   M_{inv}[b-jet + (all-non-b-jets)] =  \nonumber\\  
 =\sqrt{(P_{b-jet} +P_{all- non-b-jets})^{2}},
\end{eqnarray}
  which is constructed as the modulus of the vectorial sum of
    the 4-momentum
     $P_{b-jet}$    
    of the $b$-jet, plus the total
    4-momentum of  "all-non-$b$-jets" system
    stemming from the W decay
      ($P_{all-non-b-jets}= P_{jet1_{W}}
       + P_{jet2_{W}}$, as there are only two jets allowed 
    to be produced in W decay). 
    More precisely, if the signal event 
    contains a $\mu^{-}$ as the signal  muon, we 
    have to take the  $b$-jet ($\bar{b}$-jet would be in the case of
     $\mu^{+}$ signal muon) (see Fig.1). 
    This is only possible if one can discriminate between
    $b$- and $\bar{b}$-jets experimentally.  
    Methods of experimental determination of the charge 
    of the $b$-jet($\bar{b}$-jet) were
    developed in \cite{Damerell}.   
    In this paper  we do not  use any b-tagging procedure.
    The 
    PYTHIA information 
    on the quark flavour 
    is taken  for choosing the  
    $b$ and $\bar{b}$ jets.
    
    In the top case the invariant mass 
   $M_{inv}(b, 2~quarks_{W})$  of the system  composed
   of a $b$-quark and two quarks from W decay should 
   reproduce the mass of their parent top quark (see Fig.1).
   The distributions of 
   events 
   $dN^{event}/dM_{inv}/5$ GeV  expected  in each 
   bin of 5 GeV  
   versus  
   the  invariant mass   $M_{inv}(b, 2~quarks_{W})$ of 
   the parent three quarks
   as well as the invariant mass of jets produced by these
   quarks, i.e.  $M_{inv}(b-jet, JETS_{W})$, are shown
   for quark and jets levels in the plots
   {\bf a)} and {\bf b}) of Fig.5,  respectively, for 
   integrated luminosity 
   of 1000 $fb^{-1}$.
%
     
    The distributions  of Fig.5  show that the peak positions 
    at quark  level  (plot {\bf a)} )
    as well  as at  jet level  (plot {\bf b)} ), 
    practically coincide 
    to a good accuracy 
    as
    well as  with the  input value of the top quark mass 
    $M_{top}=170.9 (\pm 1.8)$ GeV.   

        \begin{figure}[!ht]
     \begin{center}
\vskip -0.5cm      
     \mbox{a) \epsfig{file = 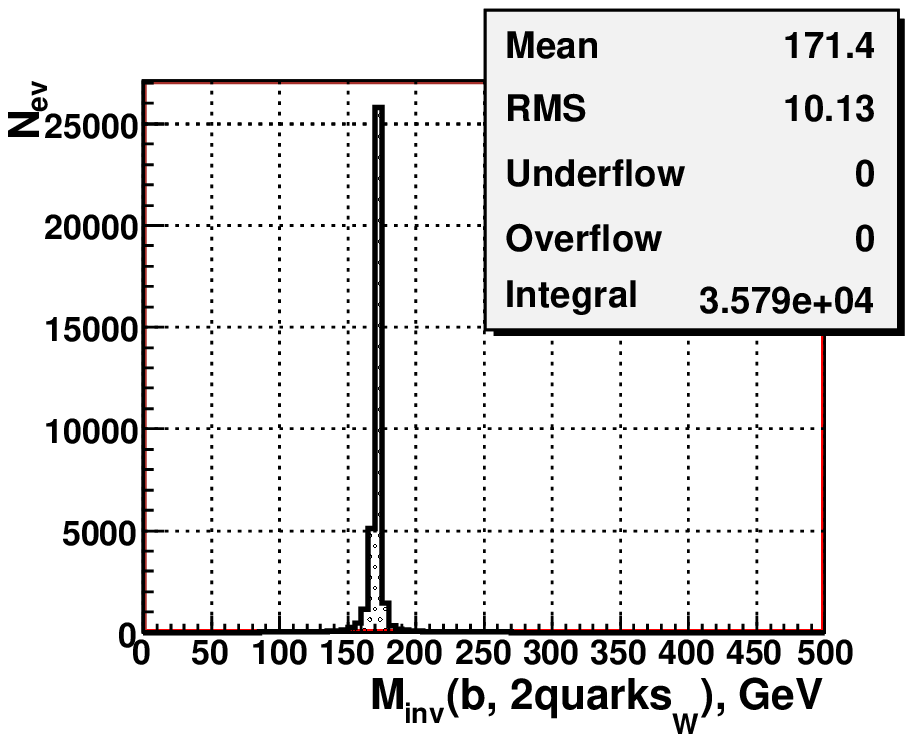 ,     
                     width=7.2cm, height=5.2cm}}     
      \mbox{b) \epsfig{file = 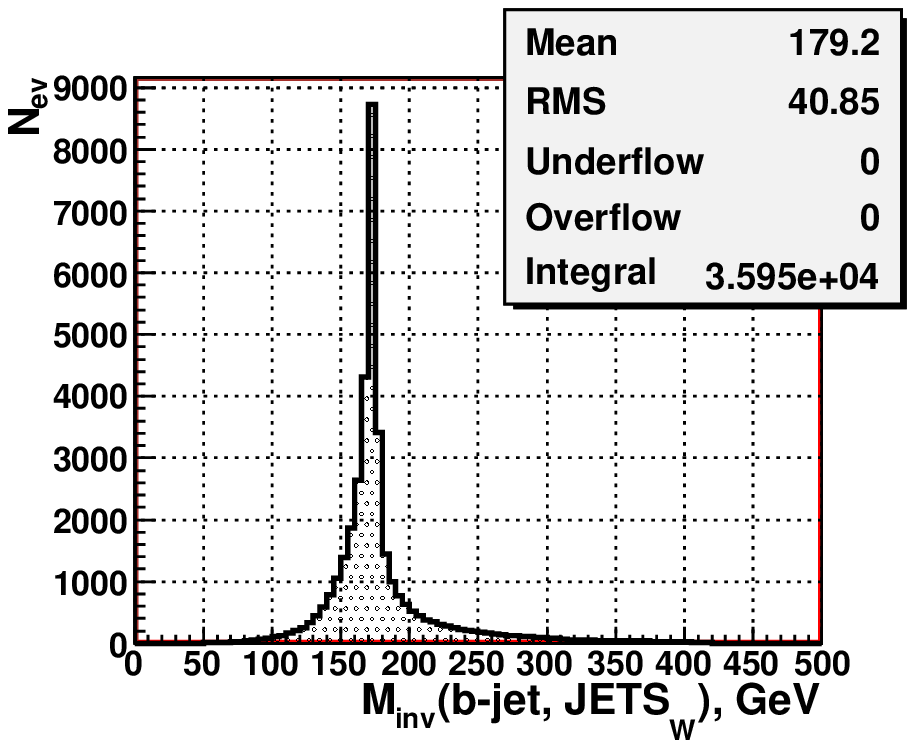,
                     width=7.2cm, height=5.2cm}}      
     \caption{\small \it The top pair  production
                    spectra of the invariant masses
	  $M_{inv}$($b$-jet,$JETS_{W}$) of the
	 "$b$-jet+(all- non-$b$-jets)" system.
	               {\bf a)}  at quark level,
	               {\bf b)}  at jet level.}
	 
     \end{center} 
\vskip -0.5cm             
    \end{figure} 
          
     It is also seen from the plot {\bf b)} of Fig.5 
     that the quark hadronisation into
     jets leads to a broadening   of   very small tails
     which are seen  in  the invariant   mass distribution at
     quark level (plot {\bf a)} of Fig.5).
    The right tail  is a bit
     lower and longer than the left one. One may
     say that the peak picture at jet level still  looks
     more or less 
     symmetric.
     The main message
     from these plots is  that the appearance of  tails 
     due to quark fragmentation into jets does not change
     the position  of the distribution peak, which allows us to 
     reconstruct the   input  top mass   both at 
     quark and jet level.           

\section{ Cuts for the reconstruction
                    of the stop mass.}
~~~~~  Based on the results of the previous 
    Section  we 
   will use  the following 
    three  cuts to  separate the signal and background: \\ 
%
%
$\bullet$ there must be at least two $b$-jets in an event: 
\begin{equation} 
     N_{b-jets} \geq 2 ; 
\end{equation}
$\bullet$ the invariant missing mass must
               be larger than $250$ GeV:
\begin{equation}   
              M_{miss}  \ge  250 ~GeV;
\end{equation}
$\bullet$ the invariant mass of  all jets  must
             be smaller than  $160$ GeV:
\begin{equation}   
               M_{inv}(All jets) \leq 160 ~GeV.
\end{equation}
	   
  They   reduce  the number of top  background
   events from 3.5*$10^4$ 
  to 
  12  and   leave 1806 stop events.   
  So, the cuts improve the 
  signal--to--background  ratio from $S/B=0.066$
  to $S/B \approx 143$ 
  losing about 
   $24\%$ of the signal stop events.
  The efficiency values for the  cuts (14)--(16)
  are calculated for $\sqrt {s}=500$ GeV. 
  We define them  as
  the summary efficiencies. It means that if 
   $\varepsilon_{1}$ 
  is the  efficiency of the first cut (14), 
   $\varepsilon_{12}$   is  the 
  efficiency of applying the first cut (14) 
  and then second cut (15).
   Analogously,  $\varepsilon_{123}$
   is the efficiency of  the successive 
   application of the cuts (14), (15) and (16).
 The following results are obtained:  \\
  
  For SIGNAL STOP events: 
%
~~~~~~~~~ $\varepsilon_{1}  =0.84$; ~~ $\varepsilon_{12}  =0.78$; ~~~~$ \varepsilon_{123}= 0.76$;

 For BACKGROUND TOP events : 
 %
 $\varepsilon_{1} =0.94$; ~~ $\varepsilon_{12}=0.001$; ~~ $\varepsilon_{123}= 3.5  \cdot 10^{-4}$. \\

  The distribution of the  invariant mass of the
  "$b$-jet+(all-non-$b$-jets)"  system in  the  case 
  of stop pair  production is shown in Fig.6.
  The plots of this figure are 
 analogous to the corresponding
   plots of Fig.5 
  using only those stop  events that have
  passed the cuts (14)--(16).  
     \begin{figure}[!ht]
     \begin{center}
\vskip -0.5cm      
     \mbox{a) \epsfig{file =  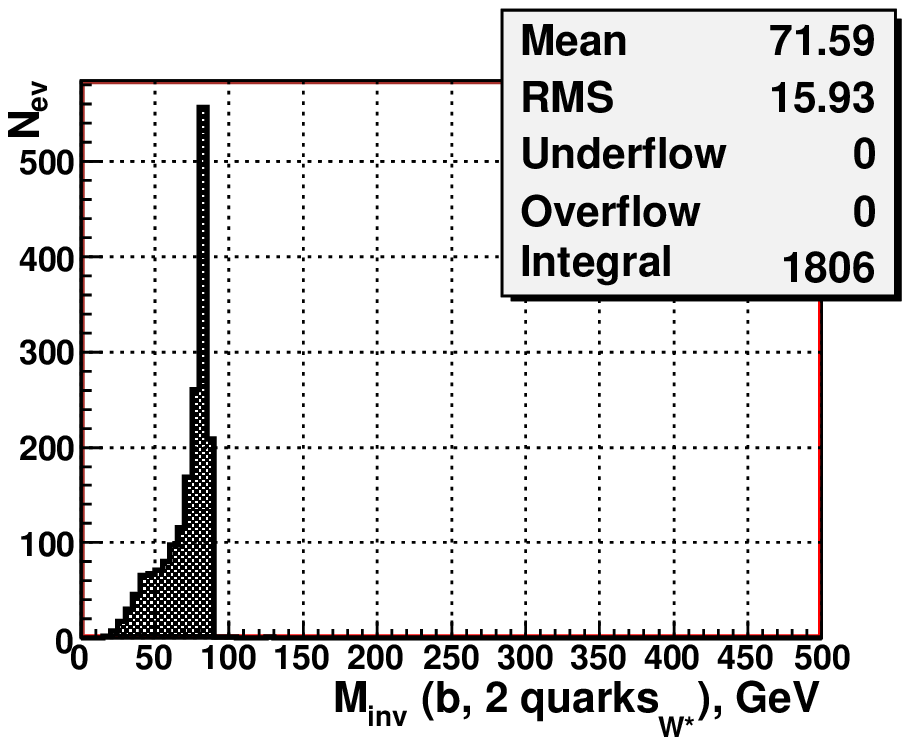,
               width=7.2cm, height=5.2cm}}  
        \mbox{b) \epsfig{file  = 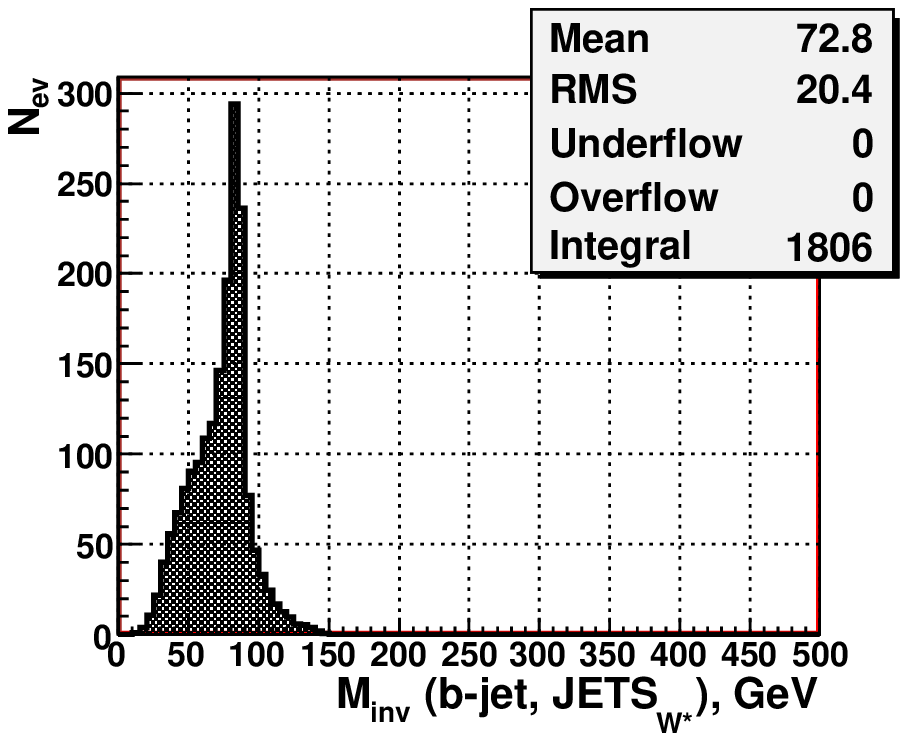 ,	
                     width=7.2cm, height=5.2cm}}  \\
     \caption{\small \it The  spectra of the  stop signal 
       events 
       after cuts versus the invariant mass
	  $M_{inv}$($b$-jet, $ JETS_{W^*}$):
	               {\bf a)} at quark level, 
	               {\bf b)} at jet level.
		       $W^*$ means virtual nature of W boson.}
     \end{center} 
\vskip -0.5cm             
    \end{figure}               

     In the stop case,  one has to take into account that 
     one of the stops decays into three jets plus a
     neutralino $\tilde \chi^{0}_{1}$. Therefore, the 
     right edge of the peak of invariant mass distribution
     of the "$b$-jet+$(JETS_{W^*})$" system
     corresponds to the mass difference 
     $M_{\widetilde{t}_1}- M_{\tilde \chi^{0}_{1}}$.      
    The position of the right edge of the peak   
    $M_{inv}$($b$-jet, $ JETS_{W^{*}})\approx$ 85~GeV     
    (plot {\bf b)} of Fig.6) 
    coincides  with the peak position seen at  quark level 
   (plot {\bf a)} of Fig.6).
   Let us note  that the observed  stability of the 
   peak position in both plots of Fig.6  is  due to
   the rather moderate  loss of the number of events
   that form this peak (they change from $\approx 550$ to 
   $\approx 280$) while passing from quark level
   to  jet level.
    Adding the  mass of the neutralino
     $M_{\tilde \chi^{0}_{1}} = 80.9$ GeV one gets 
     the reconstructed stop 
     mass $M^{reco}_{\widetilde{t}_1} \approx 166$~GeV 
     which reproduces well  (within   bin width 
     of 5 GeV) the input value 
     $M_{\widetilde{t}_1}=167.9$ GeV.

 The simulation has shown  that the 12 background
  events are mostly  distributed  in the region 
  $40 \leq M_{inv}$($b$-jet, $ JETS_{W}$)$ \leq 140$ GeV,
  which is 20 times  wider than the  5 GeV width 
  of the  peak which contains  about  280 signal stop 
  events left  after the cuts.  
   Therefore, we expect that
  in future measurements
  the contribution of a few remaining top background events
 will not influence
  the  position of this peak  which 
  allows one to reconstruct the input  value of the
  stop  mass by adding the mass of the neutralino.
  
  The origin  of the  left tail of the distribution shown 
  in the   plot  {\bf a)} of Fig.6 can be clarified  by the
   results  of the stop mass reconstruction by 
   calculating its invariant mass at quark level
   $M_{inv}$($b$, $2~quarks_{W^*}, \chi^{0}_{1})$
   as the modulus of  the sum of the  4-momenta of
   all three quarks and the neutralino (see Fig.1)
   produced in stop decay.  These results are given  in
   the plot  {\bf a)} of Fig.7 which shows a very
   precise reconstruction of the input stop mass
   at quark level withing the 5 GeV width of a peak.
\begin{figure}[!ht]
     \begin{center}
\vskip -0.5cm      
    \begin{tabular}{cc}
     \mbox{a) \epsfig{file = 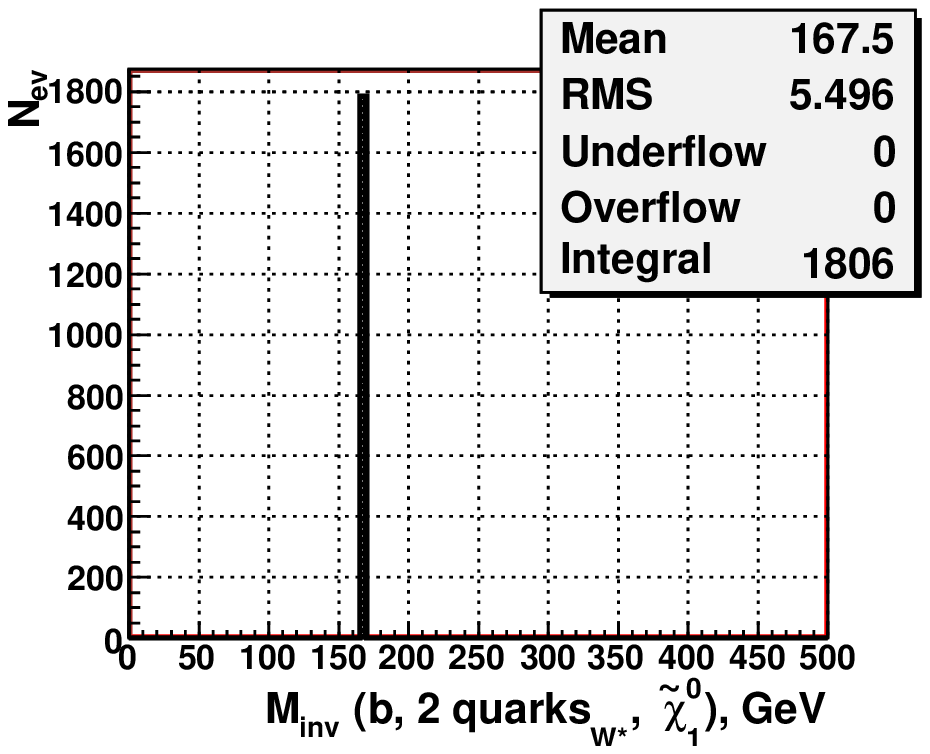,
               width=7.2cm, height=5.2cm}}     
    \mbox{b) \epsfig{file  = 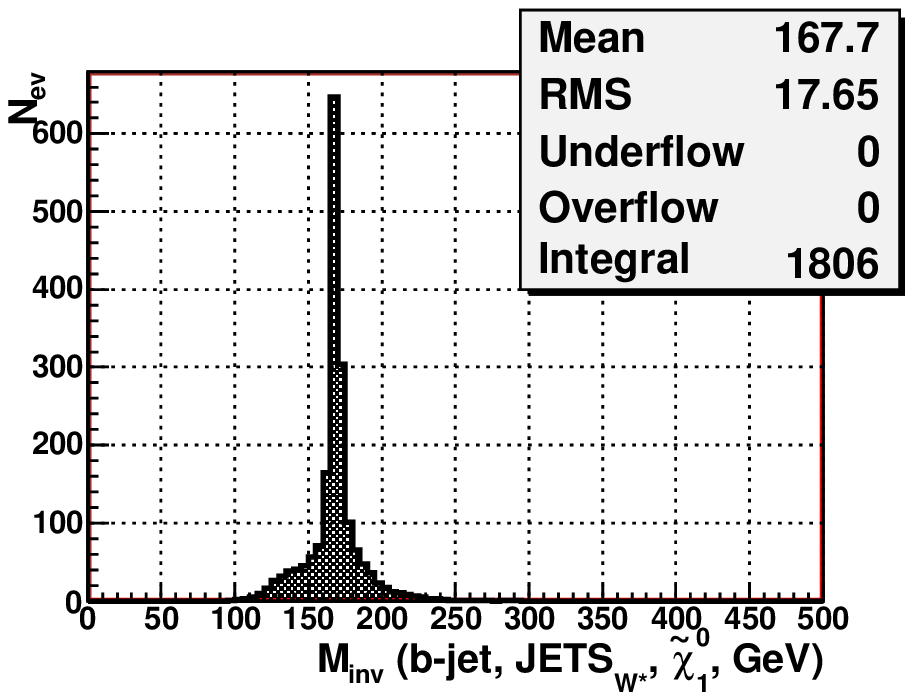,
                     width=7.2cm, height=5.2cm}}      
    \end{tabular}
     \caption{\small \it The  spectra of the  stop signal 
       events 
       after cuts versus the invariant mass
	  $M_{inv}$($b$-jet, $ JETS_{W^*}$,$\widetilde\chi^0_1$) 
	               {\bf a)} at quark level
	               {\bf b)} at jet level.}
	  
     \end{center} 
\vskip -0.5cm             
    \end{figure}     

   Comparing the plot
    {\bf a)} of Fig.6 with the plot  {\bf a)} of Fig.7
    one 
    can conclude that the long left tail as well 
    as the very small right  tail in the distribution of
    $M_{inv}$($b$, $2~quarks_{W^*})$  are  caused by
    the fact
    that the neutralino  4-momentum 
     is not included in the
    "$b$-jet+(all-non-$b$-jets)"  system.  
    Analogously, comparing  plots  {\bf b)}  and  {\bf a)} 
   of Fig.6 one may estimate  the contribution
   of the hadronisation effect (in addition to the 
   contribution of the neutralino momentum) to the 
   left tail of the distribution   {\bf b)} in Fig.6.
       
%
%
   The  influence of the
   effect of  the hadronisation of the $b$-quarks and 
   of the quarks from W decay  into jets is shown in
   the  plot  {\bf b)} of Fig.7. It is seen   that the 
   hadronisation of quarks into jets  does not 
   change the position of the stop mass peak,  
   which still  includes  the input value  
   $M_{\widetilde{t}_1}=167.9$ GeV, but
   more or less symmetrical
   and rather  suppressed short tails 
   appear around the peak
   position.
   The  shape of the peak 
   in  the stop plot {\bf b)} Fig.7  looks very similar 
   to the shape of the peak in the top  plot {\bf b)} of Fig.5,
   which demonstrates the  stability  of the 
   reconstructed top mass peak  position after 
   taking into account 
    the effect of quark
    fragmentation. 
%
%
    \subsection{ Separation of signal and  fake muons.}
%
%
  ~~~~ To select the signal stop pair production events  shown 
   in the left plot of Fig.1,
   one has to identify the
   muon from the W decay. The  distribution of the energy of
   the signal muons $ E_{sig-mu}$,
   obtained from the
   sample of  events generated without any cuts,  is shown
   in   Fig.8 {\bf a)}.
%
   \begin{figure}[!ht]
     \begin{center}
  \vskip -0.5 cm     
    \begin{tabular}{cc}
    \mbox{a) \epsfig{file =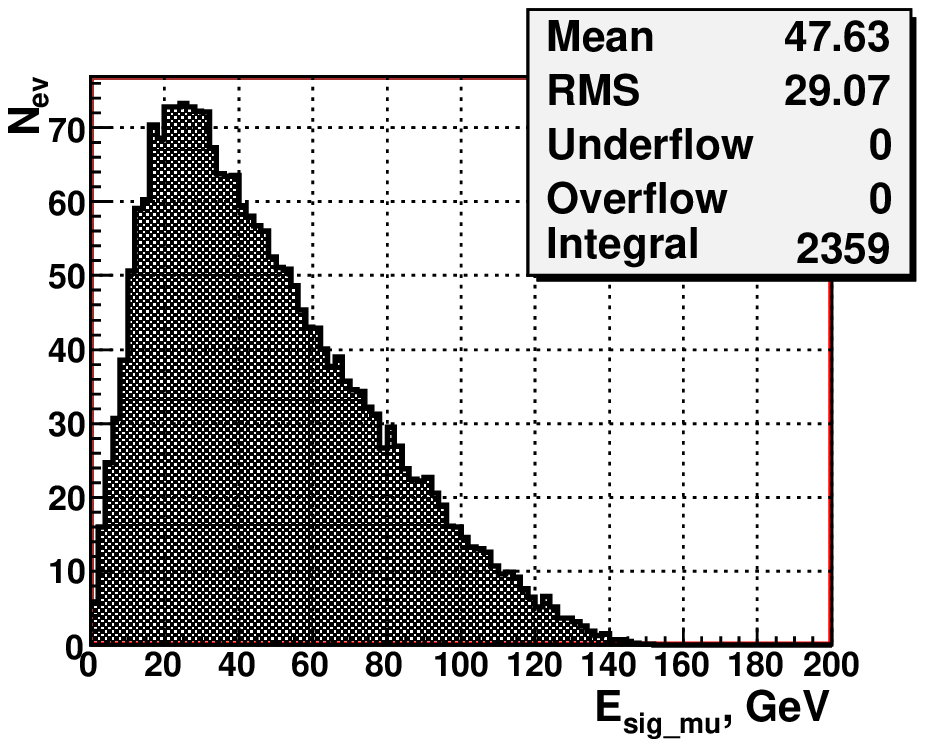, 
                                     width=7.2cm, height=5.2cm}}    
    \mbox{b) \epsfig{file = 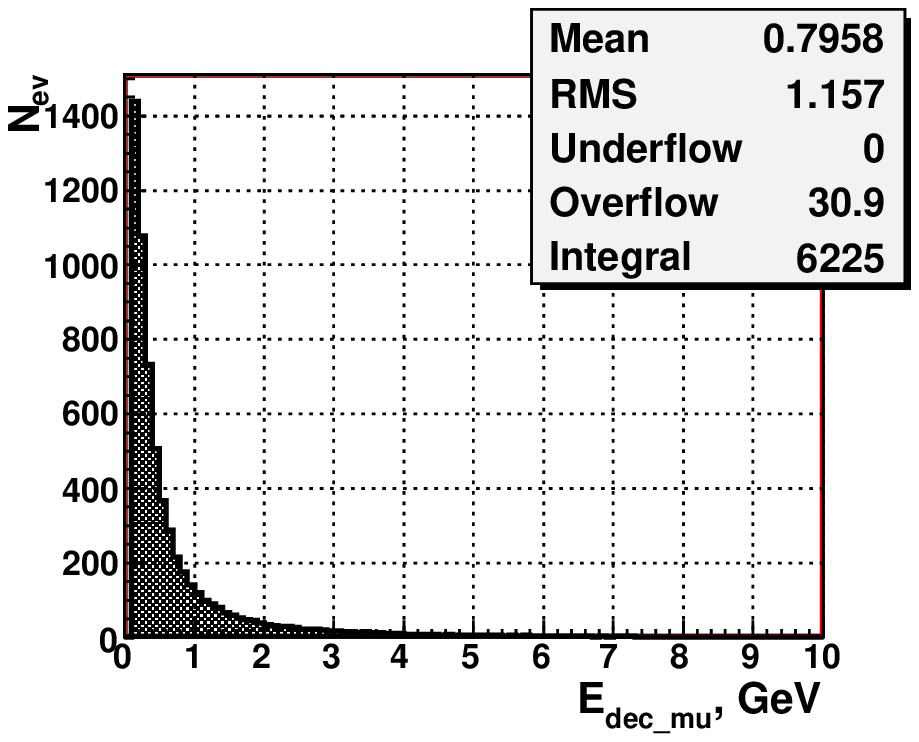,
                    width=7.2cm, height=5.2cm}}      
     \end{tabular}
     \caption{\small \it {\bf a)} Energy distribution 
                        of signal muons. 
                        {\bf b)} Energy  distributions
                        of fake muons.} 
     \end{center} 
  \vskip -0.5 cm             
     \end{figure} 
   There are, however, also 
   muons in the event coming from
   leptonic and semileptonic decays of hadrons. 
   Fig.8 {\bf b)} 
   shows the energy $E_{dec-mu}$
   of these muons  stemming from hadron 
   decays within the detector volume (for
   which we took the size 
   from  \cite{ILCRDR1}, \cite{ILCRDR2}). 
    It can be  seen that the decay muons 
    have a rather small energy  $E_{dec-mu}$. 
   Its  mean value is about $0.81$~GeV.
   The analogous spectrum  for the signal muons 
   in Fig.8 {\bf a)} shows that the  signal muons 
   have a much higher energy.
   The mean value $E_{sig-mu}$ 
   is about 50 times higher than the mean value 
   of the energy  of the decay muons. 
   One can cut off
   most low--energy decay muons rejecting those
   with  $E_{mu} \le 4$~GeV. It leads to a loss
   of about 15 signal events
   as seen from the  plot {\bf a)} of Fig.8  (the  bin 
   in this plot 
   is 2 GeV).
 
 We have also studied another way to select the signal
 muon from W decay. 
 If the axes of all four jets in the event are known,
  then in general the signal muon has the 
 largest transverse momentum
 with respect to any of  these jet axes.  
%
\subsection{ Signal and background cross sections.}
%
 ~~~~ We give in Table 1 (without cuts) and Table 2 
  (after the cuts) the cross
 sections and the  numbers of events for  stop pair
  production  and top pair production for five 
  energies  $\sqrt {s}$. To estimate the rates of 
  stop production
  at different  energies  we take 
  the  universal value of the luminosity of
  1000 fb$^{-1}$ for all energies.
  
  It is seen that the fixed cuts (14)-(16) lead to a strong
 (more than 3 orders of magnitude)   suppression
 of background top contribution and a moderate loss 
 of  signal stop events produced in the energy
 range  $400 \leq \sqrt {s} \leq 800$ GeV.  

  For the stop mass chosen, the largest
  number of signal events 
 is  expected   at  $\sqrt {s}=500$ GeV.   
   Let us note that,  according to  \cite{Damerell}, a  $50\%$
   efficiency of the separation of $b$ jets and 
   $80\%$ of the corresponding  purity can be  expected.
   It means that to get
   1806 reconstructed signal stop events  we
   will  need about 2.5 times higher statistics than
   that
   provided by  the   luminosity  1000 fb$^{-1}$
   at  $\sqrt {s}=500$ GeV. 
 
   It is worth noting  that,  as seen from
  Table 2,  even with the use 
  of the fixed parameters of the  cuts (14)--(16)
  the number of signal stop  events that can  pass
  these cuts  grows
  rapidly  with 
  the energy in the region
  $400 \leq \sqrt {s} \leq 500$ GeV where
  mass measurements will
  be done
 in the first phase.
  These  measurements may allow to enlarge the
  sample of collected signal events and to  perform 
  a precise measurement of the stop mass.
   The region  $500 \leq \sqrt {s} \leq 800$ GeV,
   which
   be available in the second phase of
   ILC operation,
   allows to gain
   a much higher stop statistics, as
   seen in 
   Table 2.
  A complete  analysis based on adjusting
  the 
  parameters of the selection cuts (15)-(16)
  for each of the
  above energy intervals
   will be presented in
   forthcoming papers.

\begin{table}[ht] 
 \vskip -0.5 cm 
\caption{ The cross sections and the number of events for stop and top pair
production before cuts.}
\begin{center} 
 \begin{tabular}{||c||c|c||c|c||c||}\hline\hline
  
  $2E_{b}=\sqrt {s} ~~[GeV]$ & 
 $\sigma^{e^{+}e^{-}}_{stop} ~~[fb]$ &
 $ N_{stop}^{events} $ & 
 $\sigma^{e^{+}e^{-}}_{top} ~~[fb]$ & 
 $ N_{top}^{events} $  &
 $ S / B $\\ \hline \hline

 350 & 0.23 & 233 & 13.76 & 13750 & 0.0169 \\
 400 & 1.34 & 1347 & 38.79 &  38740 & 0.0347 \\
 500 & 2.37 & 2378 & 35.94 &  35950 & 0.0661 \\
 800 & 1.89 & 1809 & 17.36 & 17359 & 0.1042 \\
 1000 &  1.42 & 1265 & 11.66 & 11656 & 0.1085 \\
 \hline \end{tabular}
\end{center}
 \vskip -0.5 cm  
 \end{table}
 
\begin{table}[ht] 
 \vskip -0.5 cm  
\caption{ The same as in Table 1 but after cuts.}
\begin{center}   
 \begin{tabular}{||c||c|c||c|c||c||}\hline\hline
 
  $2E_{b}=\sqrt {s} ~~[GeV]$ & 
 $\sigma^{e^{+}e^{-}}_{stop} ~~[fb]$ &
 $ N_{stop}^{events} $ & 
 $\sigma^{e^{+}e^{-}}_{top} ~~[fb]$ & 
 $ N_{top}^{events} $  &
 $ S / B $\\ \hline \hline

 350  &  0.0089 & 8    & 0                & 0   &  BKG = 0 \\
 400  &  0.52   & 521  & 2.32 * $10^{-4}$ & 0.2 &  2605    \\
 500  &  1.80   & 1806 & 2.26 * $10^{-2}$ & 12.6  &  143    \\
 800  &  0.99   & 995  & 1.08 * $10^{-2}$ & 10  &  99      \\
 1000 &  0.41   & 410  & 6.26 * $10^{-3}$ & 6   &  69      \\
 
 \hline \end{tabular}
\end{center}
\vskip -0.5 cm  
  \end{table}

\section{Conclusion.}
~~~~   We have studied  stop pair production 
   in electron-positron collisions
   for a stop mass of $167.9$~GeV 
   within the framework 
   of the  MSSM for the total energies
   $2E_{b} = \sqrt {s}$ = 350, 400, 500, 800, 1000 GeV. 
   We assume that  the 
   stop  quark   decays dominantly into a chargino and a 
   $b$ quark, $\tilde t_{1} \to b \tilde \chi_{1}^{\pm}$, 
   and the  chargino decays  into a neutralino and 
   a  W boson,
   $\tilde \chi_{1}^{\pm} \to  \tilde \chi_{1}^{0} W^{\pm}$,
   where the W boson is virtual. One of the two W's decays 
   hadronically,  $W^{+} \to q \bar q $,  the other one 
   decays leptonically,  $W^{-} \to \mu^{-} \nu$.
     
     We have performed a detailed study  based  on
    a Monte Carlo simulation with the program 
   PYTHIA6.4 for $\sqrt {s}=500$ GeV
   (at this energy we expect the highest 
   number of the signal events 
   for the chosen stop mass $M_{\tilde t_{1}}$ = 167.9 GeV
   and cuts) and the  luminosity  1000 fb$^{-1}$.  
   The program CIRCE1 is used to get the
   spectra of electron  (positron)  beams 
   taking into account the effects of beamstrahlung.
   PYTHIA6.4  is used  to simulate 
   stop pair production and decay as well as top pair
   production being the main background.

    Three cuts  (14)-(16) for the signal stop events
     have been proposed 
    to separate  signal stop events and  top 
    background events.  For  $\sqrt {s}=500$ GeV
    and  the   luminosity  1000 fb$^{-1}$ they
     give 1806 signal stop events
    with
    12 top background events.
   This is different from the more complicated 
   situation  in stop pair production at LHC 
    (see, for instance, \cite{U.Dydak}).	      
 
      We have shown that the  determination of 
   the peak position  of the distribution of the  
   invariant  mass  $M_{inv}$($b$-jet, $ JETS_{W^{*}}$) of 
   "$b$-jet+(all-non-$b$-jets)" system allows one to 
   measure  the mass of the stop quark with a good
   accuracy based on the statistics corresponding to
   the   luminosity 1000 fb$^{-1}$. 
   For this the mass of $\chi_{1}^{0}$ has to be known.
  
   As seen from the Table 2
   the measurements at other energies in the 
   regions  $400 \leq \sqrt {s} \leq 500$ GeV and 
   $500 \leq \sqrt {s} \leq 800$ GeV
   may allow to enlarge
   substancially the  number of
   selected signal stop events and to perform a precise
   measurement  of the
   mass of the scalar top quark. 
  
  In conclusion  we can say that the $e^{+}e^{-}$
  channel 
  is  well suited for the study of stop pair 
  production at  ILC. 

\section{Acknowledgements.}

 This work is supported by the JINR-BMBF project 
 and  by the "Fonds zur F$\ddot o$rderung der
 wissenschaftlichen Forschung" (FWF) of Austria, 
 project No.P18959-N16.
 The authors acknowledge support from EU
  under the MRTN-CT-2006-035505 
 and MRTN-CT-2004-503369 network programmes.
 A.B. was supported by the Spanish grants 
 SAB 2006-0072, FPA 2005-01269 and 
 FPA 2005-25348-E of the Ministero 
 de Educacion y Ciencia.


\end{document}